\documentclass[twocolumn,showpacs,superscriptaddress,10pt]{revtex4-1} 
\usepackage{amsmath,amssymb,graphicx}
\usepackage{psfrag}
\usepackage{epstopdf}
\usepackage{color}

\begin{document}

\title{Light-to-matter entanglement transfer in optomechanics}
\author{Eyob A. Sete}
\affiliation{Department of Electrical Engineering, University of California, Riverside, California 92521, USA}
\email{Corresponding author: esete@ee.ucr.edu}
\author{H. Eleuch}
\affiliation{Department of Physics, McGill University, Montreal, Canada H3A 2T8}
\author{C.H. Raymond Ooi}
\affiliation{Department of Physics, University of Malaya, Kuala Lumpur 50603, Malaysia}


\begin{abstract}
We analyze a scheme to entangle the movable mirrors of two spatially separated nanoresonators via a broadband squeezed light. We show that it is possible to transfer the EPR-type continuous-variable entanglement from the squeezed light to the mechanical motion of the movable mirrors. An optimal entanglement transfer is achieved when the nanoresonators are tuned at resonance with the vibrational frequencies of the movable mirrors and when strong optomechanical coupling is attained. Stationary entanglement of the states of the movable mirrors as strong as that of the input squeezed light can be obtained for sufficiently large optomechanical cooperativity, achievable in currently available optomechanical systems. The scheme can be used to implement long distance quantum state transfer provided that the squeezed light interacts with the nanoresonators.
\end{abstract}

\maketitle

\section{Introduction}
Quantum state transfer between two distant parties is an important and a rewarding task in quantum information processing and quantum communications. Several proposals have been put forward employing schemes based on cavity quantum electrodynamics (QED) \cite{Kua04,Li09,Seto12}. More recently quantum state transfer in quantum optomechanics, where mechanical modes are coupled to the optical modes via radiation pressure, has become a subject of interest \cite{Tia10,Cle12,Tia12,Sin12,Mcg13,Pal13}. In particular, entanglement transfer between two spatially separated cavities is appealing in quantum information. Entangling two movable mirrors of an optical ring cavity \cite{Man02}, two mirrors of two different cavities illuminated by entangled light beams \cite{Zha03}, and  two mirrors of a double-cavity set up coupled to two independent squeezed vacua \cite{Pin05} have been considered. Recently, entangling two mirrors of a ring cavity fed by two independent squeezed vacua has been proposed \cite{Hua09}. This, however, cannot be used to implement long distance entanglement transfer because the two movable mirrors belong to the same cavity.

\begin{figure*}
\includegraphics[width=15cm]{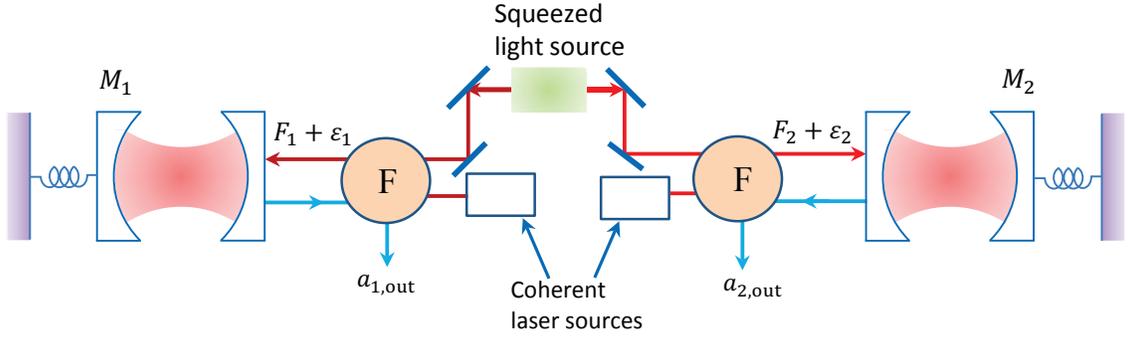}
\caption{Schematics of two nanoresonators coupled to a two-mode squeezed light from spontaneous parametric down-conversion. The output of the squeezed source is incident on the resonators as noise operators $F_{1}$ and $F_{2}$ (see text for their correlation properties). The first (second) nanoresonator movable mirror $\text{M}_{1}(\text{M}_{2})$ is coupled to the nanoresonator mode of frequency $\omega_{\text{r}_{1}}(\omega_{\text{r}_{2}})$ via radiation pressure. The nanoresonator are also driven by an external coherent laser lasers of amplitude $\varepsilon_{j}$. In the strong optomechanical coupling regime, the states of the two movable mirrors can be entangled due to the squeezed light. A Faraday isolator F is used to facilitate unidirectional coupling. The output fields $a_{1,\rm out}$ and $a_{2,\rm out}$ can be measured using the standard homodyne detection method to determine the entanglement between the mirrors. }\label{fig1}
\end{figure*}

In this work, we propose a simple model to entangle the states of two movable mirrors of spatially separated nanoresonators coupled to a common two-mode squeezed vacuum. The two-mode squeezed light, which can be generated by spontaneous parametric down-conversion, is injected into the nanoresonators as biased noise fluctuations with nonclassical correlations. The nanoresonators are also driven by two independent coherent lasers (see Fig. \ref{fig1}). The modes of the movable mirrors are coupled to their respective optical modes and to their local environments. Our analysis goes beyond the adiabatic regime \cite{Zha03} by considering the more general case of nonadiabatic regime and asymmetries between the laser drives as well as mechanical frequencies of the movable mirrors. Using parameters from a recent optomechanics experiment \cite{Gro09}, we show that the states of the two initially independent movable mirrors can be entangled in the steady state as a result of entanglement transfer from the two-mode squeezed light. More interestingly, the entanglement in the two-mode light can be totally transferred to the relative position and the total momentum of the two movable mirrors when the following conditions are met: 1) the nanoresonators are resonant with the mechanical modes, 2) the resonator field adiabatically follows the motion of the mirrors, and 3) the optomechanical coupling is sufficiently strong. We also show that the entanglement transfer is possible in the nonadiabatic regime (low mechanical quality factor), which is more closer to experimental reality. Unlike previous schemes \cite{Pin05,Hua09}, where double-or ring cavity is considered, our scheme can be used, in principle, for practical test of entanglement between two distant movable mirrors, for example, by connecting the squeezed source to the nanoresonators by an optical fiber cable. Given the recent successful experimental realization of strong optomechanical coupling \cite{Gro09}  and availability of strong squeezing up to 10 dB \cite{Vah08}, our proposal of efficient light-to-matter entanglement transfer may be realized experimentally.
\section{Model}
We consider two nanoresonators each having a movable mirror and coupled to a common two-mode squeezed vacuum reservoir, for example, from the output of the parametric down converter. One mode of the output of the squeezed vacuum is sent to the first nanoresonator and the other mode to the second nanoresonator. The movable mirror $\text{M}_{j}$ oscillates at frequency $\omega_{\text{M}_{j}}$ and interacts with the the $j$th optical mode. The $j$th nanoresonator is also pumped by external coherent drive of amplitude $\varepsilon_{j}=\sqrt{2\kappa_{j}P_{j}/\hbar \omega_{\text{L}_j}}$, where $\kappa_{j}$ is the $j$th nanoresonator damping rate, $P_{j}$ the drive pump power of the $j$th laser and  $\omega_{\text{L}_j}$ is its frequency. The schematic of our model system is depicted in Fig. \ref{fig1}. The system Hamiltonian has the form ($\hbar =1$)
\begin{align}\label{Ham}
  H&=\sum_{j=1}^{2}[\omega_{\text{M}_{j}}b^{\dag}_{j}b_{j}+\omega_{\text{r}_{j}}a^{\dag}_{j}a_{j}+\text{g}_{j}a^{\dag}_{j}a_{j}(b^{\dag}_{j}+b_{j})\notag\\
 & +(a_{j}^{\dag}\varepsilon_{j}e^{i\varphi_{j}}e^{-i\omega_{\text{L}_{j}}t}+a_{j}\varepsilon_{j}e^{-i\varphi_{j}}e^{i\omega_{\text{L}_{j}}t})],
\end{align}
where $\omega_{\text{r}_{j}}$ is the $j$th nanoresonator frequency, $\varphi_{j}$ is the phase of the $j$th input field and $\text{g}_{j}=(\omega_{\text{r}_{j}}/L_{j})\sqrt{\hbar/\text{M}_{j}\omega_{\text{M}_{j}}}$ is the single photon optomechanical coupling, which describes the coupling of the mechanical mode with the intensity of the optical mode \cite{Asp13}, where $L_{j}$ is the length of the $j$th nanoresonator and $M_{j}$ is the mass of the $j$th movable mirror; $\omega_{\text{L}_{j}}$ is the frequency of the $j$th coherent pump laser; $b_{j}$ is the annihilation operator for the $j$th mechanical mode while $a_{j}$ is the annihilation operator for the $j$th optical mode. Using the Hamiltonian \eqref{Ham}, the nonlinear quantum Langevin equations for the optical and mechanical mode variables read \cite{Tia10,Cle12,Pin05}

\begin{equation}\label{ql1}
 \dot b_{j}=-(i\omega_{\text{M}_{j}}+\frac{\gamma_{j}}{2})b_{j}-i\text{g}_{j} a_{j}^{\dag} a_{j}+\sqrt{\gamma_{j}}f_j,
\end{equation}
\begin{equation}\label{ql2}
  \dot a_{j}=-(\frac{\kappa_{j}}{2}-i\Delta_{j}) a_{j}-i\text{g}_{j}a_{j}(b_{j}^{\dag}+b_{j})-i\varepsilon_{j}e^{i\varphi_{j}}+\sqrt{\kappa_{j}} F_{j},
\end{equation}
where $\gamma_{j}$ is the $j$th movable mirror damping rate, $\Delta_{j}=\omega_{\text{L}_{j}}-\omega_{\text{r}_{j}}$ is the laser detuning, $f_{j}$ is noise operator describing the coupling of the $j$th movable mirror with its own environment while $F_{j}$ is the squeezed vacuum noise operator. Note that Eq. \eqref{ql2} is written in a frame rotating with $\omega_{\text{L}_j}$ . We assume that the mechanical baths are Markovian and have the following non zero correlation properties between their noise operators \cite{Set10,Set12}:
\begin{eqnarray}
\label{fb1}\langle f_{j}(\omega) f_{j}^{\dag}(\omega^{'}) \rangle&= &2\pi(n_{\text{th},j}+1)\delta(\omega+\omega'),\\
\label{fb2} \langle f_{j}^{\dag}(\omega)f_{j}(\omega^{'}) \rangle&=&2\pi n_{\text{th},j}\delta(\omega+\omega'),
\end{eqnarray}
where the movable mirrors are damped by the thermal baths of mean number of photons $n_{\text{th},j}=[\exp(\hbar\omega_{\text{M}_{j}}/k_{B}T_{j})-1]^{-1}$. The squeezed vacuum operators $F_{j}$ and $F_{j}^{\dag}$ have the following non vanishing correlation properties \cite{Hua09}:
\begin{eqnarray}
 \label{fc1} \langle F_{j}(\omega)F^{\dag}_{j}(\omega')\rangle &=& 2\pi (N+1)\delta(\omega+\omega'), \\
\label{fc2}  \langle F^{\dag}_{j}(\omega)F_{j}(\omega')\rangle &=& 2\pi N\delta(\omega+\omega'), \\
\label{fc3}  \langle F_{1}(\omega)F_{2}(\omega')\rangle&=& 2\pi  M \delta(\omega+\omega'-\omega_{\text{M}_{1}}-\omega_{\text{M}_{2}}),\\
\label{fc4} \langle F_{1}^{\dag}(\omega)F_{2}^{\dag}(\omega')\rangle &=&2\pi M \delta(\omega+\omega'-\omega_{\text{M}_{1}}-\omega_{\text{M}_{2}}),
\end{eqnarray}
where $N=\sinh^2 r$ and $M=\sinh r \cosh r$ with $r$ being the squeeze parameter for the squeezed vacuum light.

\section{Linearization of quantum Langevin equations}
The coupled nonlinear quantum Langevin equations [Eqs. \eqref{ql1} and \eqref{ql2}] are in general not solvable analytically. To obtain analytical solution to these equations, we adopt the following linearization scheme  \cite{Set10}. We decompose the mode operators as a sum of the steady state average and a fluctuation quantum operator as $ a_{j}= \alpha_{j}+\delta a_{j}$ and $b_{j}=\beta_{j}+\delta b_{j}$, where $\delta a_{j}$ and  $\delta b_{j}$ are operators. The mean values $\alpha_{j}$ and $\beta_{j}$ are obtained by solving
Eqs. \eqref{ql1} and \eqref{ql2} in the steady state
\begin{equation}\label{ss1}
 \alpha_{j}\equiv \langle a_{j}\rangle =\frac{-i\varepsilon_{j}e^{i\varphi_{j}}}{\kappa_{j}/2-i\Delta'_{j}},
\end{equation}
\begin{equation}\label{ss1}
   \beta_{j}\equiv\langle b_{j}\rangle =\frac{-ig_{j}|\alpha_{j}|^2}{\gamma_{j}/2+i\omega_{\text{M}_{j}}},
\end{equation}
where $\Delta'_{j}=\Delta_{j}-g_{j}(\beta_{j}+\beta^{*}_{j})$ is the effective detuning, which include the displacement of the mirrors due to the radiation pressure force. The contribution from the displacement of the movable mirrors is proportional to the intensity of the nanoresonator field, $\bar n_{j}\equiv |\alpha_{j}|^2$. In principle, we can arbitrarily choose the detunings $\Delta'_{j}$ provide that we are away from the unstable regime \cite{Set12}.

Using $a_{j}=\alpha_{j}+\delta a_{j}$ and $b_{j}=\beta_{j}+\delta b_{j}$, Eqs. \eqref{ql1} and \eqref{ql2} can be written as

\begin{eqnarray}\label{fl1}
  \delta \dot b_{j}&=&-(i\omega_{\text{M}_{j}}+\frac{\gamma_{j}}{2})\delta b_{j}+\mathcal{G}_{j}(\delta a_{j}-\delta a_{j}^{\dag})+\sqrt{\gamma_{j}}f_{j},\\
\label{fl2}
  \delta \dot a_{j}&=&-(\frac{\kappa_{j}}{2}-i\Delta'_{j})\delta  a_{j}-\mathcal{G}_{j}(\delta b_{j}^{\dag}+\delta b_{j})+\sqrt{\kappa_{j}}F_{j},
\end{eqnarray}
where $\mathcal{G}_{j}\equiv g_{j}|\alpha_{j}|=g_{j}\sqrt{\bar n_{j}}$ is the many-photon optomechanical coupling. Since the phase of the coherent drives can be arbitrary, for convenience we have chosen the phase of the input field to be $\varphi_{j}=-\arctan({2\Delta'_{j}/\kappa_{j}})$ so that $\alpha_{j}=-i|\alpha_{j}|$. Notice that the linearized equations \eqref{fl1} and \eqref{fl2} can be described by an effective Hamiltonian ($\hbar =1$)
\begin{align}\label{Heff}
 \mathcal{H}&=\sum_{j=1}^{2}\Big[\omega_{\text{M}_j}\delta b_{j}^{\dag}\delta b_{j}-\Delta_{j}'\delta a_{j}^{\dag}\delta a_{j}\notag\\
 &+i\mathcal{G}_{j}(\delta a_{j}-\delta a_{j}^{\dag})(\delta b_{j}+\delta b_{j}^{\dag})\Big]
\end{align}
with a new effective many-photon optomechanical coupling $\mathcal{G}_{j}$, which is stronger than the single photon coupling $g_{j}$ by a factor of $\sqrt{\bar n_{j}}$. The effective Hamiltonian \eqref{Heff} describes two different processes depending on the choice of the laser detuning $\Delta'_{j}$ \cite{Asp13}. Here we want emphasize that $\omega_{\rm M_{j}}\gg \gamma_{j}$ and $\Delta_{j}\gg \kappa_{j}$ so that we can apply the rotating wave approximation. The latter condition is the case when the resonators are strongly off-resonant with the laser fields. When $\Delta'_{j}=-\omega_{\text{M}_{j}}$, within the rotating wave approximation, the interaction Hamiltonian reduces to $\mathcal{H}_{I}=-i\sum_{j=1}^{2}\mathcal{G}_{j}(\delta a_{j}\delta b_{j}^{\dag}-\delta a_{j}^{\dag}\delta b_{j})$, which is relevant for quantum state transfer \cite{Tia10,Cle12} and cooling (transferring  of all thermal phonons into cold photon mode) \cite{Cha11}. In quantum optics, it is referred to as a 'beam-splitter' interaction. Whereas, when $\Delta'_{j}=+\omega_{\text{M}_{j}}$ (in rotating wave approximation), the interaction Hamiltonian takes a simple form $\mathcal{H}_{I}=-i\sum_{j=1}^{2}\mathcal{G}_{j}(\delta a_{j}\delta b_{j}-\delta a_{j}^{\dag}\delta b_{j}^{\dag})$, which describes parametric amplification interaction and can be used for efficient generation of optomechanical squeezing and entanglement. In this work, we are interested in quantum state transfer and hence choose $\Delta'_{j}=-\omega_{\text{M}_{j}}$. Thus, for $\Delta'_{j}=-\omega_{\text{M}_{j}}$ and in a frame rotating with frequency $\omega_{\text{M}_{j}}$ (neglecting the fast oscillating terms), one gets
\begin{eqnarray}\label{flu1}
  \delta \dot{\tilde b}_{j}&=&-\frac{\gamma_{j}}{2}\delta \tilde b_{j}+\mathcal{G}_{j}\delta \tilde a_{j}+\sqrt{\gamma_{j}}\tilde f_{j},\\
\label{flu2}
  \delta \dot {\tilde a}_{j}&=&-\frac{\kappa_{j}}{2} \delta \tilde a_{j}-\mathcal{G}_{j}\delta \tilde b_{j}+\sqrt{\kappa_{j}}\tilde F_{j},
\end{eqnarray}
where we have introduced a notation for operators: $\tilde o=o \exp(i\omega_{\text{M}_{j}}t)$.

In the following section we use these equations to analyze the entanglement of the states of the movable mirrors via entanglement transfer.

\section{Entanglement analysis}
In order to investigate the entanglement between the states of the movable mirrors of the two spatially separated nanoresonators, we introduce two EPR-type quadrature operators for the mirrors, namely their relative position $X$ and the total momentum $Y$: $X=X_{1}-X_{2}$ and $Y=Y_{1}+Y_{2}$, where $X_{l}=(\delta \tilde b_{l}+\delta \tilde b^{\dag}_{l})/\sqrt{2}$ and $Y_{l}=i(\delta \tilde b^{\dag}_{l}-\delta \tilde b_{l})/\sqrt{2}$. We apply entanglement criterion \cite{Dun00} for continuous variables which is sufficient for nonGaussian states, and sufficient and necessary for Gaussian states. According to this criterion, the states of the movable mirrors are entangled if
 \begin{equation}\label{crit}
   \Delta X^2+\Delta Y^2<2.
 \end{equation}
Thus for maximally entangled states (EPR or Bell states) the total variance becomes 0, while for separable states the sum of the variances will be equal or greater than 2.
\subsection{Adiabatic regime}
 An optimal quantum state transfer (in this case from the two-mode squeezed vacuum to the mechanical motion of the mirrors) is achieved when the nanoresonator fields adiabatically follow the mirrors, $\kappa_{j}\gg \gamma_{j},\mathcal{G}_{j}$ \cite{Pin05}, which is the case for mirrors with high-Q mechanical factor and weak effective optomechanical coupling. (In fact the condition $\kappa_{j}\gg \gamma_{j}$ can also be expressed as $\omega_{\rm r_{j}}\gg \omega_{\rm M_{j}}(Q_{\rm r_{j}}/Q_{\rm M_{j}})$.) Inserting the steady state solution of \eqref{flu2} into \eqref{flu1}, we obtain equations describing the dynamics of the movable mirrors
\begin{equation}\label{mm1}
    \delta \dot {\tilde b}_{j}=-\frac{\Gamma_{j}}{2}\delta \tilde b_{j}+\sqrt{\Gamma_{a_{j}}}\tilde F_{j}+\sqrt{\gamma_{j}}\tilde f_{j},
 \end{equation}
where $\Gamma_{j}=\Gamma_{a_{j}}+\gamma_{j}$ with $\Gamma_{a_{j}}=4\mathcal{G}_{j}^2/\kappa_{j}$ being the effective damping rate induced by the radiation pressure \cite{Met04}.

First, let us consider the variance of the relative position of the two mirrors  $\Delta X^2$, which can be expressed as $\Delta X^2=\langle X^2\rangle-\langle X\rangle^2$. Since the noise operators corresponding to the two-mode squeezed vacuum $F_{j}$ as well as the movable mirrors baths $f_{j}$ have zero mean values, it is easy to show that $\langle X\rangle=0$. Therefore, $\Delta X^2=\langle X_{1}^2\rangle+\langle X_{2}^2\rangle-\langle X_{1}X_{2}\rangle-\langle X_{2}X_{1}\rangle$.
To evaluate these correlations, it is more convenient to work in frequency domain. To this end, the Fourier transform of Eqs. \eqref{mm1} yields
\begin{align}
  \delta \tilde b_{j}(\omega) = \frac{\sqrt{\Gamma_{a_{j}}}\tilde F_{j}(\omega)+\sqrt{\gamma_{j}}\tilde f_{j}(\omega)}{\Gamma_{j}/2+i\omega}.
 \end{align}
The expectation value of the position $X_{1}$ of the first movable mirror can be expressed as
\begin{equation}\label{DX1}
  \langle X^2_{1}\rangle=\frac{1}{4\pi^2}\int_{-\infty}^{\infty}\int_{-\infty}^{\infty}d\omega d\omega' e^{i(\omega+\omega' )t}\langle X_{1}(\omega)X_{1}(\omega')\rangle.
\end{equation}
Using the correlation properties of the noise operators [Eqs. \eqref{fb1}-\eqref{fc4}], we obtain
\begin{align}\label{DX2}
  \langle X^2_{1}\rangle=\frac{1}{2}(2N+1)\frac{\Gamma_{a_{1}}}{\Gamma_{1}}+\frac{\gamma_{1}}{2\Gamma_{1}}(2n_{\text{th},1}+1).
\end{align}
Similarly, it is easy to show that
\begin{eqnarray}
 \label{DX3}\langle X^2_{2}\rangle&=&\frac{1}{2}(2N+1)\frac{\Gamma_{a_{2}}}{\Gamma_{2}}+\frac{\gamma_{2}}{2\Gamma_{2}}(2n_{\text{th},2}+1).\\
  \label{DX4} \langle X_{1}X_{2}\rangle&=&\langle X_{2}X_{1}\rangle =\frac{2\sqrt{\Gamma_{a_{1}}\Gamma_{a_{2}}}}{\Gamma_{1}+\Gamma_{2}}M.
\end{eqnarray}
Therefore, using Eqs. \eqref{DX2}-\eqref{DX4}, the variance of the relative position of the movable mirrors becomes
\begin{align}\label{DX}
  \Delta X^2&=\frac{1}{2}(2N+1)\left(\frac{\Gamma_{a_{1}}}{\Gamma_{1}}+\frac{\Gamma_{a_{2}}}{\Gamma_{2}}\right)-\frac{4\sqrt{\Gamma_{a_{1}}\Gamma_{a_{2}}}}{\Gamma_{1}+\Gamma_{2}}M\notag\\
  &+\frac{\gamma_{1}}{2\Gamma_{1}}(2n_{\text{th},1}+1)+\frac{\gamma_{2}}{2\Gamma_{2}}(2n_{\text{th},2}+1).
\end{align}
It is easy to show that the variance of the total momentum of the movable mirrors is the same as that of $X$, i.e., $\Delta X^2=\Delta Y^2$. Thus, the sum of the variances of the relative position and total momentum of the movable mirrors is given by

\begin{figure}[t]
\includegraphics[width=6cm]{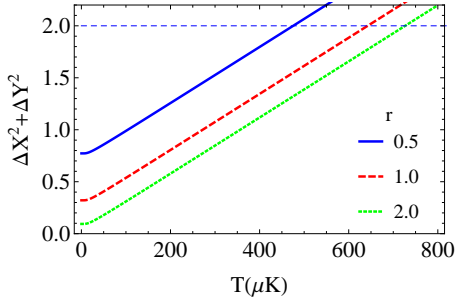}
  \caption{Plots of the sum of variances  $\Delta X^2+\Delta Y^2$ vs bath temperature $T$ of the movable mirrors for drive laser power $P=10~\text{mW}$ and frequency $\omega_{L}=2\pi\times 2.82\times 10^{14}~\text{Hz} (\lambda=1064~\text{nm})$, mass of the movable mirrors $\text{M}_{1}=\text{M}_{2}=145~\text{ng}$, frequency of the nanoresonator $\omega_{r}=2\pi\times 5.26\times 10^{14}~\text{Hz}$, length of the cavity $L=125~\text{mm}$, the mechanical motion damping rate $\gamma=2\pi\times 140~\text{Hz}$, $\omega_{\text{M}}=2\pi\times 947\times 10^{3}~\text{Hz}$, nanoresonator damping rate $\kappa=2\pi\times 215\times 10^{3}~\text{Hz}$, and for different values of the squeezing parameter $r$: 0.5 (blue solid curve), 1.0 (red dashed curve), and 2.0 (green dotted curve). The blue dashed line represents $\Delta X^2+\Delta Y^2=2$.}\label{fig2}
\end{figure}

\begin{align}\label{DX}
  &\Delta X^2+\Delta Y^2=\frac{\gamma_{1}}{\Gamma_{1}}(2n_{\text{th},1}+1)+\frac{\gamma_{2}}{\Gamma_{2}}(2n_{\text{th},2}+1)\notag\\
  &+(2N+1)\left(\frac{\Gamma_{a_{1}}}{\Gamma_{1}}+\frac{\Gamma_{a_{2}}}{\Gamma_{2}}\right)
  -\frac{8\sqrt{\Gamma_{a_{1}}\Gamma_{a_{2}}}}{\Gamma_{1}+\Gamma_{2}}M.
\end{align}

\subsubsection{Identical nanoresonators}

To elucidate the physics of light-to-matter entanglement transfer, we first consider a simplified case of identical nanoresonators coupled to two-mode squeezed vacuum. We also assume the external laser drives to have the same strength and the thermal baths of the two movable mirrors to be at the same temperature ($n_{\text{th},1}=n_{\text{th},2}=n_{\text{th}}$). To this end, setting $\Gamma_{1}=\Gamma_{2}=\Gamma$, $\Gamma_{a_{1}}=\Gamma_{a_{2}}=\Gamma_{a}$, $\text{M}_{1}=\text{M}_{2}$, $\omega_{r}=\omega_{r_{1}}=\omega_{r_{2}}$, $\omega_{\rm M}=\omega_{M_{1}}=\omega_{M_{2}}$, $\kappa=\kappa_{1}=\kappa_{2}$, and $\gamma_{1}=\gamma_{2}=\gamma$, and using the relation $N=\sinh ^2 r, M=\sinh r \cosh r$, the variance of the relative position \eqref{DX} takes a simple form
\begin{align}\label{Sim}
  \Delta X^2+  \Delta Y^2&=\frac{2\Gamma_{a}}{\gamma+\Gamma_{a}}e^{-2r}+\frac{2\gamma}{\gamma+\Gamma_{a}}(2n_{\text{th}}+1)\notag\\
  &=\frac{8e^{-2r}\mathcal{G}^2/\gamma\kappa+2+4n_{\text{th}}}{4\mathcal{G}^2/\gamma\kappa+1}\notag\\
  &=\frac{2\mathcal{C}}{\mathcal{C}+1}e^{-2r}+\frac{2(1+2n_{\text{th}})}{\mathcal{C}+1},
\end{align}
where $\mathcal{C}=4\mathcal{G}^2/\gamma\kappa=4\bar ng^2/\gamma\kappa$ is the optomechanical cooperativity \cite{Pur13}. In the absence of the two-mode squeezed vacuum reservoir $r=0$, Eq. \eqref{Sim} reduces to $\Delta X^2+\Delta Y^2=2+4n_{\text{th}}/(\mathcal{C}+1)$, which is always greater than 2, indicating the mechanical motion of the two mirrors cannot be entangled without the squeezed vacuum. This is because the motion of the mirrors are initially uncorrelated and their interaction via vacuum does not create correlations. In the limit $\mathcal{C}\gg 1$ (a weaker condition \cite{Cle08} for strong coupling regime), the sum of the variances can be approximated by $\Delta X^2+\Delta Y^2\approx2\exp(-2r)+4n_{\text{th}}/\mathcal{C}$. Therefore, when $4n_{\text{th}}/\mathcal{C}<1$, which can be achieved for sufficiently large number of photons in the nanoresonator, the sum of the variances can be less than 2 when
\begin{align}
r>\frac{1}{2}\ln [1/(1-2n_{\text{th}}/\mathcal{C})],
\end{align}
indicating transfer of the quantum fluctuations of the input fields to the motion of the movable mirrors. This can be interpreted as entanglement transfer from light to mechanical motion. The interesting aspect is that this quantum state transfer scheme can, in principle, be extended to long distance state transfer if the two nanoresonators are kept far apart but connected by, for example, an optical fiber cable to the output of the two-mode squeezed vacuum. Obviously, the entanglement between the mirrors would degrade when the distance between the resonators is increased owing to the decrease in degree of squeezing as a result of environmental couplings.
Recently, similar transfer scheme from light to matter has been proposed \cite{Kua04,Dan04,Dan044}.

For realistic estimation of the entanglement between the movable mirrors, we use parameters from recent experiment \cite{Gro09}: laser frequency $\omega_{L}=2\pi\times 2.82\times 10^{14}~\text{Hz} (\lambda=1064 ~\text{nm})$, $\omega_{r}=2\pi\times 5.64\times 10^{14}~\text{Hz}~(\omega_{r}=2\omega_{L})$, $M_{1}=M_{2}=145~\text{ng}$, $L=25~\text{mm}$, $\kappa=2\pi\times 215\times 10^{3}\text{Hz}$, $\gamma=2\pi\times 140~\text{Hz}$, $\omega_{M}=2\pi\times 947 \times 10^{3}~\text{Hz}$. In Fig. \ref{fig2}, we plot the sum of the variances of $X$ and $Y$ as a function the temperature of the thermal bath of the movable mirrors. This figure shows that the movable mirrors are entangled when the nanoresonators are fed with squeezed light. Notice that based on the definition of the quadrature operators $X$ and $Y$, an optomechanical quadrature squeezing \cite{Set12,Set14,Pur13} is achieved when $\Delta X^2<1$ or $\Delta Y^2<1$. This implies that whenever there is optomechanical squeezing, the two movable mirrors are always entangled. This shows a direct relationship between optomechanical squeezing and entanglement of the mechanical modes of the movable mirrors.

\begin{figure}[t]
\includegraphics[width=6cm]{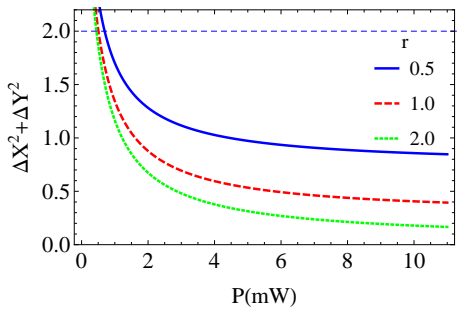}
  \caption{Plots of the sum of variances  $\Delta X^2+\Delta Y^2$ vs drive pump power for thermal bath temperature $T=50~\mu K$ of the movable mirrors, $\omega_{r}=2\pi\times 2.82\times 10^{14}~\text{Hz}$, and for different values of the squeezing parameter $r$: 0.5 (blue solid curve), 1.0 (red dashed curve), and 2.0 (green dotted curve). All other parameters as the same as in Fig. \ref{fig2}. The blue dashed line represents $\Delta X^2+\Delta Y^2=2$.}\label{fig3}
\end{figure}

It is also interesting to see the dependence of the mirror-mirror entanglement on the pump laser power strength. Figure \ref{fig3} shows that for a given squeeze parameter $r$ and the thermal bath temperature $T$ of the movable mirrors, there exists a minimum pump power strength for which the movable mirrors are entangled. The minimum power required to observe mirror-mirror entanglement can be derived from \eqref{Sim} by imposing the condition that $\Delta X^2+\Delta Y^2<2$, which yields
\begin{equation}
\mathcal{C}> \frac{2n_{\text{th}}}{1-\exp(-2r)}.
\end{equation}
Using the explicit form of $\mathcal{G}$ in $\mathcal{C}=4\mathcal{G}^2/\gamma\kappa$, we then obtain ($r\neq0)$
\begin{equation}\label{min}
  P> \frac{\alpha}{(1-e^{-2r})(\exp[\hbar\omega_{\text{M}}/k_{B}T]-1)},
\end{equation}
where $\alpha \equiv\gamma \omega M_{1} L^2 \omega_{\text{M}}[(\kappa/2)^2+\omega_{\text{M}}^2]/2\omega_{r}^2$ is a factor which can be fixed at the beginning of the experiment (note here that $M_{1}=M_{2})$. It is easy to see from  \eqref{min} that for a given thermal bath temperature $T$ of the movable mirrors, increasing $r$ decreases the minimum power required to achieve entanglement between the mirrors.

\begin{figure}[t]
\includegraphics[width=6cm]{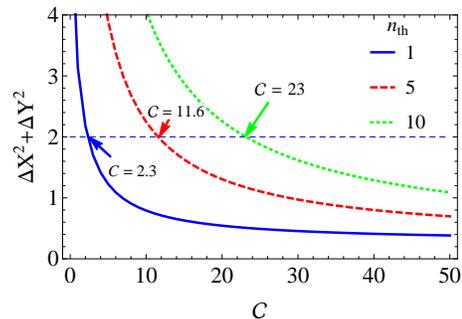}
  \caption{Plots of the sum of variances $\Delta X^2+\Delta Y^2$ vs the optomechanical cooperativity $\mathcal{C}$ for squeeze parameter $r=1$ and for various values of the thermal bath photon numbers: $n_{\text{th}}=1 (T=62.2\mu \text{K})$ (green dotted curve), $n_{\text{th}}=5~(T=236~\mu \text{K})$ (red dashed curve), and $n_{\text{th}}=10 ~(T=452~\mu \text{K})$ (blue solid curve).}\label{figey}
\end{figure}

When the number of thermal bath photons increases, the minimum value of the cooperativity parameter for which the entanglement occurs increases. In the weak coupling regime, where the optomechanical cooperatively is much less than one, $\mathcal{C}\ll1$, the sum of the variances \eqref{Sim} that characterize the entanglement can be approximated by $\Delta X^2+\Delta Y^2\approx 2+2\mathcal{C}e^{-2r}+4n$. This is always greater than 2 independent of the degree of squeezing of the input field, indicating no quantum state transfer from the squeezed light to the mechanical motion of the movable mirrors, and hence the mirrors remain unentangled. Figure \ref{figey} shows the plot of the entanglement measure vs the optomechanical cooperativity as a function of the thermal bath photon numbers. For $r=1.0$ and $n=1.0~(62.2 ~\mu\text{K})$ the motion of the two mirrors are not entangled up to $\mathcal{C}=2n_{\text{th}}[1-\exp(-2r)]^{-1}\approx 2.3$.

\subsubsection{Effect of asymmetric coherent drives and mechanical frequencies}

\begin{figure}[t]
\includegraphics[width=6cm]{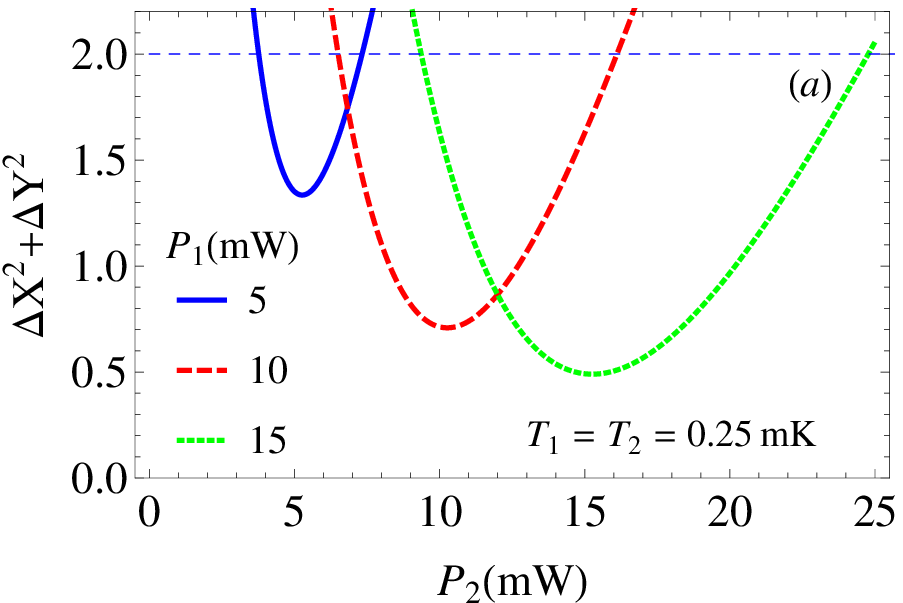}
\includegraphics[width=6cm]{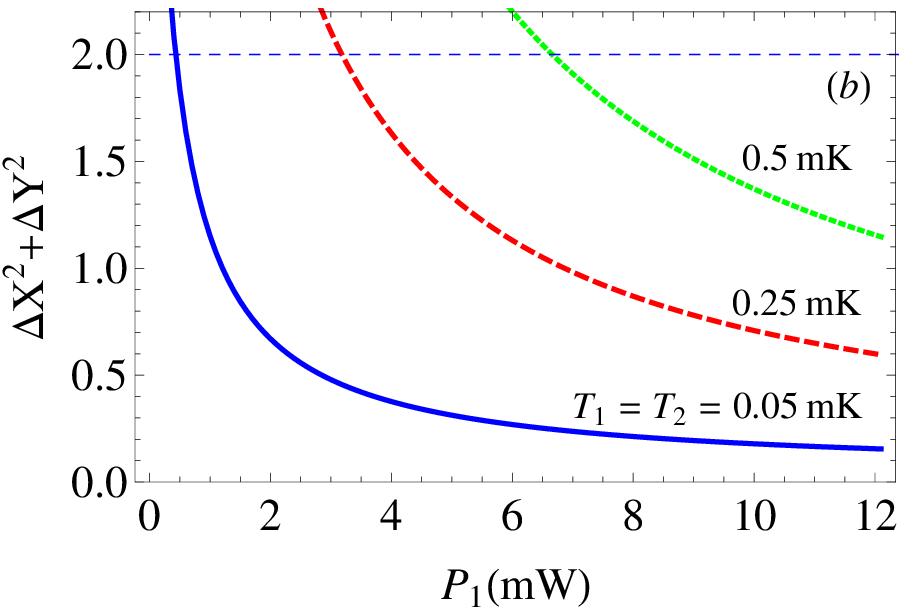}
\caption{(a)$\Delta X^2+\Delta Y^2$ vs the input drive power $P_{2}$ of the second nanoresonator and for various values of the input drive power $P_{1}$ of the first nanoresonator and assuming the same thermal bath temperatures of the movable mirrors $T_{1}=T_{2}=0.25~\text{mK}$ and squeeze parameter $r=2.0$. (b) $\Delta X^2+\Delta Y^2$ vs the input drive power of the first nanoresonator optimized over the input power of the second nanoresonator and for different values of $T_{1}$ and $T_{2}$ and squeeze parameter $r=2.0$. All other parameters are the same as in Fig. \ref{fig2}. The blue dashed line in both figures represents $\Delta X^2+\Delta Y^2=2$.}\label{fig5}
\end{figure}

We next analyze the effect of the asymmetries in the strength of coherent drives and in the vibrational frequencies of the movable mirrors. Figure \ref{fig5}a illustrates that for a constant thermal bath temperatures $T_{1}=T_{2}=0.25~\text{mK}$ of the movable mirrors and squeeze parameter $r=2.0$, there exist input laser powers $P_{1}$ and $P_{2}$, where $\Delta X^2+\Delta Y^2$  is minimum or the entanglement is the strongest. It turns out that for identical nanoresonators, strong entanglement is achieved when $P_{1}=P_{2}$.  Notice also that the width of the entanglement region is mainly determined by the input power: the higher the input powers, the wider the width of entanglement region becomes.Figure \ref{fig5}b shows optimized $\Delta X^2+\Delta Y^2$ over the input power $P_{2}$ for a given $P_{1}$ for different values of the thermal bath temperatures $T_{1}$ and $ T_{2}$. As expected the entanglement degrades as the thermal bath temperatures of the mirrors increase and the entanglement persists at higher temperatures for sufficiently strong pump power strength (see green-dotted curve for $T_{1}=T_{2}=0.5\text{mK}$.)

\begin{figure}[t]
\includegraphics[width=6cm]{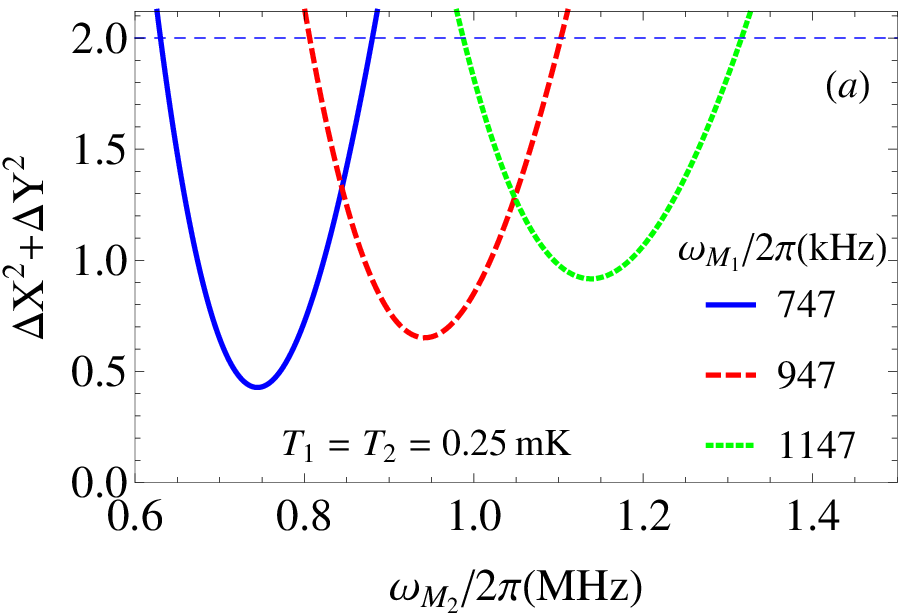}
\includegraphics[width=6cm]{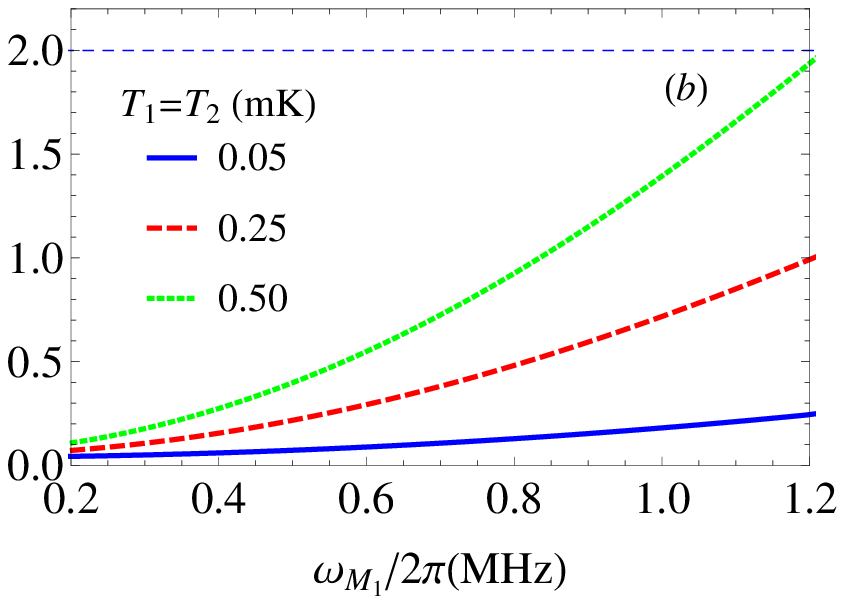}
\caption{(a)$\Delta X^2+\Delta Y^2$ vs the vibrational frequency $\omega_{\rm M_{2}}$ of the second nanoresonator and for various values of the vibrational frequency $\omega_{\rm M_{1}}$ of the first nanoresonator and assuming the input laser powers $P_{1}=P_{2}=11~\text{mW}$ and squeeze parameter $r=2.0$. (b) $\Delta X^2+\Delta Y^2$ vs the vibrational frequency $\omega_{\rm M_{1}}$ of the first nanoresonator optimized over $\omega_{\rm M_{2}}$ and for different values of $T_{1}$ and $T_{2}$ and squeeze parameter $r=2.0$.  All other parameters are the same as in Fig. \ref{fig2}. The blue dashed line in both figures represents $\Delta X^2+\Delta Y^2=2$. }\label{fig6}
\end{figure}

Tuning the frequencies of the movable mirrors also affects the degree of the mirror-mirror entanglement. As shown in Fig. \ref{fig6}a, for a fixed temperatures of the thermal bath of the movable mirrors $T_{1}=T_{2}=0.25\text{mK}$ and squeeze parameter $r=2.0$ and drive powers $P_{1}=P_{2}=11\text{mW}$ and the frequency $\omega_{\rm M_{1}}$ of the first movable mirror, there exists a frequency $\omega_{\rm M_{2}}$ of the second movable mirror for which the entanglement is maximum. The smaller $\omega_{\rm M_{1}}$ is, the stronger the entanglement becomes. The optimum entanglement decreases with increasing frequency $\omega_{\rm M_{1}}$ of the first movable mirror and eventually disappears at sufficiently large  $\omega_{\rm M_{1}}$  and relatively high temperatures (see Fig. \ref{fig6}b.)

\subsection{Nonadiabatic regime}
So far we have discussed the mirror-mirror entanglement induced by the squeezed light in the adiabatic regime ($\kappa_{j}\gg \gamma_{j},\mathcal{G}_{j}$). We next derive a condition for entanglement valid for both adiabatic and nonadiabatic regimes. We also study the field-field entanglement in the regime where the two mirrors are entangled.

The dynamics of the movable mirrors in the nonadiabatic regime is described by the coupled equations \eqref{flu1} and \eqref{flu2}. Solving the Fourier transforms of these equations yields

\begin{equation}\label{nad}
  \delta \tilde b_{j}=\frac{\kappa_{j}/2+i\omega}{d_{j}(\omega)}\sqrt{\gamma_{j}} \tilde f_{j}+\frac{\mathcal{G}_{j}}{d_{j}(\omega)}\sqrt{\kappa_{j}}\tilde F_{j},
\end{equation}
where $d_{j}(\omega)=\mathcal{G}_{j}^2+(\gamma_{j}/2+i\omega)(\kappa_{j}/2+i\omega)$. Thus using \eqref{nad} and the properties of the noise operators \eqref{fb1}-\eqref{fc4}, the sum of the variances of the relative position $X$ and total momentum $Y$ of the movable mirrors (for identical nanoresonators) is found to be
\begin{equation}\label{nv}
  \Delta X^2+\Delta Y^2=\frac{2\mathcal{C}}{\mathcal{C}+1}\frac{\kappa~ e^{-2r}}{\kappa+\gamma}+\frac{2(2n_{\text{th}}+1)}{\mathcal{C}+1}\big[1+\frac{\mathcal{C}\gamma}{\kappa+\gamma}\big].
\end{equation}
We immediately see that for $\kappa\gg\gamma, \mathcal{G}$, Eq. \eqref{nv} reduces to the expression \eqref{Sim} derived in the adiabatic approximation. In general, for the dissipation rate of the movable mirrors $\gamma_{j}$ comparable to the resonator decay $\kappa_{j}$, the expression \eqref{nv} can be significantly different from \eqref{Sim}.

\begin{figure}
\includegraphics[width=6cm]{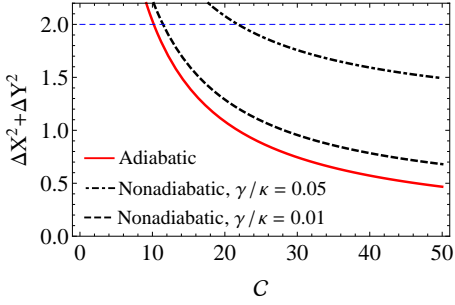}
\caption{Plots of the sum of the variance of the quadrature operators $X$, $Y$ for the mirror versus the optomechanical cooperativity parameter in the adiabatic regime [\eqref{Sim}] (red solid curve) and in the nonadiabatic regime [Eq. \eqref{nv}] for $\gamma/\kappa=0.01$ (black dashed curve) and $0.05$ (black dotdashed curve). Here we used $n_{\text{th}}=5$ and squeeze parameter $r=2$. The blue dashed line in both figures represents $\Delta X^2+\Delta Y^2=2$.\label{nonad}}
\end{figure}

In Fig. \ref{nonad} we present a comparison showing the entanglement transfer in the adiabatic and nonadiabatic regimes. The main difference comes from the mechanical dissipation rate $\gamma$. Since the adiabatic approximation assumes negligible mechanical dissipation rate, the transfer is more efficient than the non adiabatic case. This however is an ideal situation, which requires very high mechanical quality factor. In general, for low mechanical quality factor the mechanical dissipation can be significant, leading to a less efficient entanglement transfer. As can be noted from Fig. \ref{nonad}, the mirror-mirror entanglement diminishes when the normalized mechanical dissipation rate $\gamma/\kappa$ increases from $0.01$ to $0.05$. We note that when the dissipation rate increases, large cooperativity (strong coupling) is required to observe the mirror-mirror entanglement.

To gain insight into the transfer of entanglement from the squeezed light to the motion of the mirrors, it is important to study the entanglement between the optical modes of the nanoresonators. This can be analyzed by introducing two EPR-type quadrature operators $x=x_{1}-x_{2}$ and $y=y_{1}+y_{2}$, where $x_{l}=(\delta \tilde a_{l}+\delta \tilde a_{l}^{\dag})/\sqrt{2}$ and $y_{l}=i(\delta \tilde a^{\dag}_{l}-\delta \tilde a_{l})/\sqrt{2}$. The optical modes of the nanoresonators are entangled if
\begin{equation}\label{ff}
  \Delta x^2+\Delta y^2<2.
\end{equation}
Solving the Fourier transforms of Eqs. \eqref{ql1} and \eqref{ql2}, we obtain
\begin{align}\label{Q1}
  \delta \tilde a_{j}(\omega)=-\frac{\mathcal{G}_{j}}{d_{j}(\omega)}\sqrt{\gamma_{1}}\tilde f_{j}(\omega)+\frac{\gamma_{j}/2+i\omega}{d_{j}(\omega)}\sqrt{\kappa_{j}}\tilde F_{j},
\end{align}
where $d_{j}(\omega)=\mathcal{G}_{j}^2+(\kappa_{j}/2+i\omega)(\gamma_{j}/2+i\omega)$. The sum of the variances of $x$ and $y$ for identical nanoresonators reads
\begin{align}\label{vx}
  \Delta x^2+\Delta y^2&=\frac{2\mathcal{C}(2n_{\text{th}}+1)}{\mathcal{C}+1}\frac{\gamma}{\gamma+\kappa}\notag\\
  &+2\left(\frac{\kappa}{\kappa+\gamma}+\frac{1}{1+\mathcal{C}}\frac{\gamma}{\gamma+\kappa}\right)e^{-2r}
\end{align}
which for the case $\gamma/\kappa\ll1$ and strong coupling regime ($\mathcal{C}\gg1$) reduces to
\begin{align}\label{vx-limit}
  \Delta x^2+\Delta y^2\approx 2(2n_{\text{th}}+1)\frac{\gamma}{\gamma+\kappa}+2e^{-2r}.
\end{align}
\begin{figure}
\includegraphics[width=6cm]{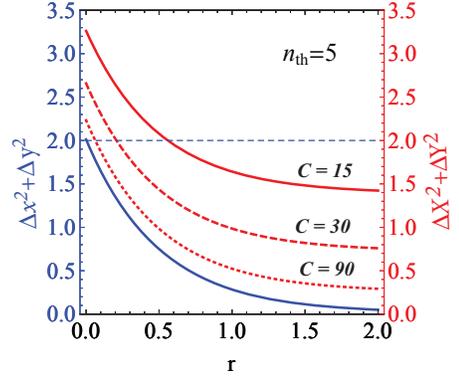}
\caption{Plots of the sum of the variance of the quadrature operators $X$ and $Y$ for the mirror $\Delta X^2+\Delta Y^2$[Eq. \eqref{nv}] (red curves with different $\mathcal{C}$ values) and the quadrature operators $x$, $y$ for the field $\Delta x^2+\Delta y^2$[Eq. \eqref{vx}] (blue solid curve) vs the squeeze parameter $r$ for different values of the optomechanical cooperativity parameter $\mathcal{C}= 15$ (red solid curve), 30 (red dashed curve), and 90 (red dotted curve). Here we used $\gamma/\kappa=6.5\times 10^{-4}$, $n_{\text{th}}=5$. The blue dashed line shows $\Delta x^2+\Delta y^2=\Delta X^2+\Delta Y^2=2$ below which the stationary states of the movable mirrors as well as the nanoresonator modes are entangled.}\label{fig9}
\end{figure}
We note from \eqref{vx-limit} that in the strong coupling regime, the field-field entanglement is mainly determined by the thermal bath temperature and squeeze parameter, not on the value of $\mathcal{C}$. For experimental parameter in Ref. \cite{Man02} we have $\gamma/\kappa=6.5\times 10^{-4}$ and assuming the thermal bath mean photon number $n_{\text{th}}=5$, the field-field entanglement is insensitive to the increase of the cooperativity while the entanglement of the states of the movable mirrors increases as the optomechanical coupling becomes stronger or $\mathcal{C}$ increases (Fig. \ref{fig9}). It is interesting to see that for sufficiently strong coupling (large values of cooperativity, $\mathcal{C}$), the entanglement between the states of the movable mirrors can be as strong as that of the squeezed light. Therefore, in addition to choosing the mechanical frequency to be $\Delta'=-\omega_{\text{M}}$ and adiabatic approximation ($\kappa\gg \gamma,\mathcal{G}$), it is imperative to attain strong coupling regime to achieve the maximum entanglement between the states of the movable mirrors.

Experimentally, the entanglement between the states of the movable mirrors can be measured by monitoring the phase and amplitude \cite{Man02} of the transmitted field via the method of homodyne detection, in which the signal is brought into interference with a local oscillator that serves as phase reference. For other variants of optical measurement schemes see Ref. \cite{Asp13}. With the availability \cite{Vah08} of strong squeezing sources up to 10 dB squeezing ($90\%$) below the standard quantum limit, our proposal can be realized experimentally.

\section{Conclusion}
In summary, we have analyzed a scheme to entangle the vibrational modes of two independent movable mirrors and spatially separated nanoresonators via two-mode squeezed light. We showed that in the regime of strong coupling $\mathcal{C}\gg 1 (4\mathcal{G}^2\gg \kappa\gamma)$ and when the nanoresonator field adiabatically follows the motion of the mirrors, the quantum fluctuations of the two-mode squeezed light is transferred to the motion of the movable mirrors, creating stationary entanglement between the vibrational modes of the movable mirrors. It turns out that an entanglement of the states of the movable mirrors as strong as the entanglement of the two-mode squeezed light can be achieved for sufficiently large optomechanical cooperativity $\mathcal{C}$ or equivalently for sufficiently strong optomechanical coupling. We also considered a less stringent condition--nonadiabatic regime which is more realistic than the adiabatic approximation and still obtained entanglement transfer from the two-mode light to the movable mirrors. Given the recent successful experimental realization of strong optomechanical coupling \cite{Gro09} and well-developed method of homodyne measurement, our proposal for efficient light-to-matter entanglement transfer may be realized experimentally.

\begin{acknowledgements}
EAS acknowledges financial support from the
Office of the Director of National Intelligence (ODNI), Intelligence
Advanced Research Projects Activity (IARPA), through the Army
Research Office Grant No. W911NF-10-1-0334. All statements of fact,
opinion or conclusions contained herein are those of the authors and
should not be construed as representing the official views or
policies of IARPA, the ODNI, or the U.S. Government. He also
acknowledge support from the ARO MURI Grant No. W911NF-11-1-0268.
\end{acknowledgements}

 \end{document}